\documentclass[11pt]{article}
\usepackage{fullpage}
\usepackage{latexsym}
\usepackage{mathptmx}
\usepackage[scaled]{helvet}
\usepackage[dvips]{graphicx}
\usepackage{amsmath, amssymb, amsthm, amsfonts}

\newcommand{\eat}[1]{}
\newcommand{\junk}[1]{}

\newcommand{\pr}{\mathbb{P}}
\newcommand{\ex}{\mathbb{E}}

\newtheorem{fact}{Fact}

\newtheorem{lemma}{Lemma}
\newtheorem{theorem}{Theorem}

\newtheorem{definition}{Definition}

\title{A General Framework for Graph Sparsification}
\author{Ramesh Hariharan\\Strand Life Sciences \and
Debmalya Panigrahi\\CSAIL, MIT}
\date{}

\begin{document}

\maketitle

\begin{abstract}
Given a weighted graph $G$ and an error parameter $\epsilon > 0$, 
the {\em graph sparsification} problem requires
sampling edges in $G$ and giving the sampled 
edges appropriate weights to obtain a sparse graph $G_{\epsilon}$ with the following
property: the weight of every cut in $G_{\epsilon}$ is within a factor of
$(1\pm \epsilon)$ of the 
weight of the corresponding cut in $G$. 
Bencz\'ur and Karger~\cite{BenczurK96} showed how to obtain $G_{\epsilon}$ 
with $O(n\log n/\epsilon^2)$ edges in time $O(m\log^3 n)$ for weighted graphs
and $O(m\log^2 n)$ for unweighted graphs using a combinatorial approach based on strong connectivity.
Spielman {\em et al}~\cite{SpielmanS08} showed  how to obtain $G_{\epsilon}$ 
with $O(n\log n/\epsilon^2)$ edges in time $O(m\log^c n)$ for some (large) constant $c$
using an algebraic approach based on effective resistances.
Our contributions are as below (all for weighted graphs $G$ with $n$ vertices
and $m$ edges having polynomial-sized weights, unless otherwise stated):
\begin{itemize}
\item Bencz\'ur and Karger~\cite{BenczurK96} conjectured that 
using standard connectivity instead of
strong connectivity for sampling would simplify the result substantially, 
and posed this as an open question. 
In this correspondence, we resolve this question by showing that sampling 
using standard connectivities also preserves cut weights and yields a 
$G_{\epsilon}$ with  $O(n\log^2 n/\epsilon^2)$ edges.
\item We provide a very simple strictly linear time algorithm (i.e. $O(m)$ time) for 
graph sparsification that yields a $G_{\epsilon}$ with  $O(n\log^2 n/\epsilon^2)$ edges.
\item We provide another algorithm for 
graph sparsification that yields a $G_{\epsilon}$ with  $O(n\log n/\epsilon^2)$ edges in
$O(m\log^2 n)$ time (for unweighted graphs, this reduces to $O(m\log n)$ time).
\item Combining the above two results,
we obtain the fastest known algorithm for obtaining a $G_{\epsilon}$ with 
$O(n\log n/\epsilon^2)$ edges;
this algorithm runs in time $O(m + n\log^4 n/\epsilon^2)$ whereas the previous best bound 
is $O(m\log^3 n)$.
\item If $G$ has arbitrary edge weights, we give an $O(m\log^2 n)$-time
algorithm that yields a $G_{\epsilon}$ containing 
$O(n\log^2 n/\epsilon^2)$ edges. The previous best bound is $O(m\log^3 n)$
time for a $G_{\epsilon}$ with $O(n\log n/\epsilon^2)$ edges.
\item Most importantly, we provide a generic framework that sets out sufficient
conditions for any particular sampling scheme to result in good sparsifiers; 
all the above results can
be obtained by simple instantiations of this framework, as can known results on
sampling by strong connectivity and sampling by effective resistances\footnote{
with a $G_{\epsilon}$ that is slightly denser than the 
best-known result for the effective resistance case.}.
\end{itemize}
Our algorithms are Monte-Carlo, i.e. work with high probability, as are
all efficient algorithms for graph sparsification.

A key ingredient of our proofs is a generalization of 
bounds on the number of small cuts in an undirected graph
due to Karger~\cite{Karger93}; 
this generalization might be of independent interest.

\end{abstract}

\newpage

\section{Introduction}
\label{sec:intro}

A {\em cut} of an undirected graph is a partition of its
vertices into two disjoint sets. The {\em weight} of a
cut is the sum of weights of the edges crossing the cut, 
i.e. edges having one endpoint each in the two vertex 
subsets of the partition. For unweighted graphs, each edge
is assumed to have unit weight. Cuts play an important
role in many problems in graphs: e.g., the 
maximum flow between a pair of vertices is equal to the
minimum weight cut separating them.

A {\em skeleton} $G'$ of an undirected graph $G$ is a subgraph 
of $G$ on the same set of vertices where each edge in $G'$
can have an arbitrary weight. In a series of results, 
Karger~\cite{Karger94a, Karger94b} showed that an appropriately 
weighted sparse skeleton generated by {\em random sampling}
of edges approximately preserves the weight of 
{\em every} cut in an undirected graph. This series of results
culminated in a seminal work by Bencz\'ur and 
Karger~\cite{BenczurK96} that showed the following theorem.
Throughout this paper, 
for any undirected graph $G$ and any $\epsilon\in (0, 1]$, 
$(1\pm \epsilon)G$ is the set of all appropriately weighted 
subgraphs of $G$ where the weight of every cut in the subgraph
is within a factor of $(1\pm \epsilon)$ of the weight of the 
corresponding cut in $G$.
\begin{theorem}[Bencz\'ur-Karger~\cite{BenczurK96}]
\label{thm:sampling}
For any undirected graph $G$ with $m$ edges and
$n$ vertices, and for any error parameter $\epsilon\in (0, 1]$, 
there exists a skeleton $G_{\epsilon}$ containing 
$O(\frac{n\log n}{\epsilon^2})$ edges such that 
$G_{\epsilon}\in (1\pm\epsilon)G$ 
with high probability.\footnote{We say that a property holds
{\em with high probability} (or whp) for a graph on $n$ vertices
if its failure probability can be bounded by the inverse of
a fixed polynomial in $n$.} 
Further, such a skeleton can be found in $O(m\log^2 n)$ time
if $G$ is unweighted and $O(m\log^3 n)$ time otherwise.
\end{theorem}
\noindent 
Besides its combinatorial ramifications, the importance of this 
result stems from its use as a pre-processing step in several graph 
algorithms, e.g. to obtain an $\tilde{O}(n^{3/2}+m)$-time algorithm for approximate 
maximum flow using the $\tilde{O}(m\sqrt{m})$-time algorithm for exact maxflow 
due to Goldberg and Rao~\cite{GoldbergR98}; and more recently, 
$\tilde{O}(n^{3/2}+m)$-time algorithms for approximate sparsest 
cut~\cite{KhandekarRV09, Sherman09}. 


Subsequent to Bencz\'ur and Karger's work,
Spielman and Teng~\cite{SpielmanT04, SpielmanT06} extended their results to 
preserving all quadratic forms, of which cuts are a special case;
however, the size of the skeleton constructed was $O(n\log^c n)$
for some large constant $c$. Spielman and 
Srivastava~\cite{SpielmanS08} improved this result by constructing skeletons of
size $O(\frac{n\log n}{\epsilon^2})$ in $O(m\log^{O(1)} n)$ time, 
while continuing to preserve all quadratic 
forms. Recently, this result was further improved by Batson {\em et al}~\cite{BatsonSS09} 
who gave a deterministic algorithm for constructing skeletons of 
size $O(\frac{n}{\epsilon^2})$. While their result is optimal in terms of
the size of the skeleton constructed, the time complexity of their
algorithm is $O(\frac{mn^3}{\epsilon^2})$, rendering it somewhat useless 
in terms of applications. 

Bencz\'ur and Karger~\cite{BenczurK96}, and
Spielman {\em et al}~\cite{SpielmanT04, SpielmanT06, SpielmanS08, BatsonSS09}
use contrasting techniques to obtain their respective results; the former 
use combinatorial graph techniques while the latter use algebraic graph 
techniques. In each case, the goal is to obtain a probability value $p_e$ for each edge
$e$ so that sampling each edge $e$ independently with probability $p_e$ and giving
each sampled edge $e$ a weight $1/p_e$ yields $G_{\epsilon}\in (1\pm\epsilon)G$.
Bencz\'ur and Karger~\cite{BenczurK96} choose $p_e$ inversely proportional to the 
{\em strong connectivity} of $e$ while 
Spielman {\em et al}~\cite{SpielmanT04, SpielmanT06, SpielmanS08, BatsonSS09}
choose $p_e$ proportional to the {\em effective resistance} of $e$ (both concepts are
defined below).

\begin{definition}
The {\em strong connectivity} of an edge $(u, v)$ in an undirected graph
$G$ is the maximum value of $k$ such that there is an induced subgraph 
$G'$ of $G$ containing both $u$ and $v$, and every cut in $G'$ has weight
at least $k$.
\end{definition}

\begin{definition}
The {\em effective resistance} of an edge $(u, v)$ in an undirected 
graph $G$ is the effective electrical resistance between $u$ and $v$ 
if each edge in $G$ is replaced by an electrical resistor between its
endpoints whose electrical resistance is equal to the weight of the 
edge. 
\end{definition}



\subsection{Our Results} 

We obtain the following results.

\noindent\paragraph{The Generic Framework.}
We provide a general proof framework as follows. 
For any given sampling scheme (i.e., assignment to the $p_e$'s),
we show that if this assignment satisfies two sufficient conditions, then
the sampling scheme results in good sparsifiers.
All of the results stated below are then simple instantiations of
the above framework, i.e. we show that the sufficient conditions hold.
The resulting algorithms are also much simpler than 
those in \cite{BenczurK96} or in \cite{SpielmanS08, BatsonSS09}.

\eat{
Thus, we are able to 
 unifies various sampling techniques 
in graphs---using standard connectivity (new result), 
strong connectivity~\cite{BenczurK96} and effective
resistance~\cite{SpielmanS08, BatsonSS09} of edges. In addition, this
generic scheme allows us to design extremely simple 
algorithms to construct such skeleton graphs. For all ranges
of weights, our algorithms are more efficient than those due
to Bencz\'ur and Karger~\cite{BenczurK96}, and if the edge 
weights are bounded by a polynomial in the number of vertices, 
our algorithms are more efficient than any previous algorithm
and attain optimal time complexity in terms of the number of 
edges in the input graph. The skeletons we construct are slightly
larger than those due to Bencz\'ur and Karger~\cite{BenczurK96}, and 
Spielman and Srivastava~\cite{SpielmanS08}, but one can run
a previously known (slower) algorithm on the skeleton that we construct 
using our (faster) algorithm on the input graph 
to obtain the best of both worlds, i.e. sparsity
of the skeleton provided by the previous results and the greater 
efficiency of our algorithm. (Note that the previous algorithms 
run on a sparse skeleton and not on the input graph; so their
larger time complexity does not affect the running time of the 
overall two-step algorithm.) In the 
course of proving these results, we also extend combinatorial theorems
for counting small cuts, thereby providing more insight into the 
connectivity structure of undirected graphs.
}

\noindent\paragraph{Faster Algorithms.}

Our first result is an efficient algorithm for constructing a sparse 
skeleton. 
\begin{theorem}
\label{thm:algo}
Suppose $G$ is an undirected graph with $n$ vertices and $m$ edges. 
Then, for any fixed $\epsilon\in (0, 1]$, 
there is an efficient algorithm for finding a skeleton $G_{\epsilon}$ 
of $G$ having
$O(\frac{n\log n}{\epsilon^2})$ edges in expectation such that 
$G_{\epsilon}\in (1\pm\epsilon)G$ whp.
The time complexity of the algorithm is 
$O(m + n\log^4 n/\epsilon^2)$ if the weights of all edges are bounded by a fixed 
polynomial in $n$ (including all unweighted graphs).
\end{theorem}
\noindent
This is the first sampling algorithm that runs in time strictly linear
in $m$; all previous algorithms 
had a time bound of at least $O(m\log^2 n)$ for unweighted graphs,
and $O(m\log^3 n)$ for weighted graphs. This algorithm improves the
time complexity of several problems, where creating a graph sparsifier
in the first step. We mention some of these applications.
\begin{itemize}
\item This yields an $O(m) + \tilde{O}(n^{3/2}/\epsilon^3)$-time algorithm
for finding the $\epsilon$-approximate 
maximum flow between two vertices of an undirected graph
using the exact maxflow algorithm in~\cite{GoldbergR98}. 
The previous best
algorithm had a running time of $O(m\log^3 n) + \tilde{O}(n^{3/2}/\epsilon^3)$.
\item This yields an $O(m) + \tilde{O}(n^{3/2})$-time algorithm for
finding an $O(\log n)$-approximate sparsest cut~\cite{KhandekarRV09, Sherman09},
and an $O(m) + \tilde{O}(n^{3/2+\delta})$-time algorithm for
finding an $O(\sqrt{\log n})$-approximate sparsest cut for any 
constant $\delta$~\cite{Sherman09}. 
The previous best algorithms had running time of 
$O(m\log^3 n) + \tilde{O}(n^{3/2})$ and $O(m\log^3 n) + \tilde{O}(n^{3/2+\delta})$
respectively.
\end{itemize}
The sampling algorithm in Theorem~\ref{thm:algo} 
is obtained by composing two different algorithms
described below. The first algorithm is fast but 
generates a slightly denser skeleton. The second (slower) algorithm then operates
on this skeleton to obtain a smaller skeleton.

\begin{theorem}
\label{thm:algo2}
Suppose $G$ is an undirected graph with $n$ vertices and $m$ edges. 
Then, for any fixed $\epsilon \in (0, 1]$, 
there is an efficient algorithm for finding a skeleton $G_{\epsilon}$ 
of $G$ having
$O(\frac{n\log^2 n}{\epsilon^2})$ edges in expectation such that
$G_{\epsilon}\in (1\pm \epsilon)G$ whp.
The time complexity of the algorithm is
$O(m)$ if the weights of all edges are bounded by a fixed 
polynomial in $n$ (including all unweighted graphs), and $O(m\log^2 n)$ 
if the edges have arbitrary weights.
\end{theorem}


\begin{theorem}
\label{thm:algo2-1}
Suppose $G$ is an undirected graph with $n$ vertices and $m$ edges. 
Then, for any fixed $\epsilon \in (0, 1]$, 
there is an algorithm for finding a skeleton $G_{\epsilon}$ 
of $G$ having
$O(\frac{n\log n}{\epsilon^2})$ edges in expectation such that
$G_{\epsilon}\in (1\pm \epsilon)G$ whp.
The time complexity of the algorithm is
$O(m\log n)$ for unweighted graphs and $O(m\log^2 n)$ if the weights of all edges are bounded by a fixed 
polynomial in $n$ (including all unweighted graphs).
\end{theorem}

\noindent\paragraph{Sampling by Standard Connectivity, Effective
Resistances and Strong Connectivity.}
In proving Theorem~\ref{thm:sampling}, the authors had to use strong
connectivity because the more natural notion of {\em standard connectivities} seemed
to pose complications. 
\begin{definition}
The {\em standard connectivity}, or simply {\em connectivity}, of an edge 
$(u, v)$ in an undirected graph $G$ is the maximum flow between $u$ and $v$
in $G$.
\end{definition}
\noindent
The authors conjectured that using standard connectivity instead of
strong connectivity for sampling would simplify the result substantially,
and posed this as their main open question. In this correspondence,
we resolve this question by showing that sampling
using standard connectivities also preserves cut weights. 

\begin{theorem}
\label{thm:standard}
Suppose $G$ is an undirected graph on $n$ vertices. 
For any fixed $\epsilon\in (0, 1]$, let $G_{\epsilon}$ be a skeleton of 
$G$ formed by sampling edge $e$ in $G$ with 
probability\footnote{$\ln n = \log_e n; \lg n = \log_2 n.$}
$p_e = \min(\frac{96(3+\lg n)\ln n}{0.38 k_e \epsilon^2}, 1)$, where $k_e$
is the standard connectivity of edge $e$ in $G$. If selected in the 
sample, edge $e$ is given a weight of $1/p_e$ in the skeleton. Then, 
$G_{\epsilon}$
has $O(\frac{n\log^2 n}{\epsilon^2})$ edges in expectation and
$G_{\epsilon}\in (1\pm\epsilon)G$ whp.
\end{theorem}
\noindent Observe that the size of the skeleton constructed
using standard connectivity has an extra $\log n$ factor 
compared to that constructed using strong connectivity. We conjecture that
this factor can indeed be removed by more careful analysis.

We show that exactly the same proof as above  holds if we replace standard connectivity
with {\em effective resistance} of an edge. Thus, we show that 
sampling edges using effective resistances also produces a sparse
skeleton that approximately preserves all cut weights, a result 
independently obtained by Spielman and Srivastava recently for the 
larger class of all quadratic forms (cuts are a special type of 
quadratic forms) with a tighter bound on the size of the 
skeleton~\cite{SpielmanS08}. Our result, though weaker, has a 
much simpler proof.

We also show that the results obtained in~\cite{BenczurK96} using
strong connectivity can be obtained as a simple instantiation
of our general sampling framework.

\noindent\paragraph{Generalizations of Cut Counting.}

The {\em edge connectivity} of an undirected graph is the minimum 
weight of a cut in the graph.
A key ingredient in the proof of Theorem~\ref{thm:sampling} is a
celebrated theorem due to Karger~\cite{Karger93}) 
that gives tight bounds on the
number of distinct cuts of a fixed weight in an undirected graph
in terms of the ratio of the weight of the cuts to the edge 
connectivity of the graph.
\begin{theorem}[Karger~\cite{Karger93}]
\label{thm:cut-count}
For an undirected graph with edge connectivity $c$ and for any
$\alpha \geq 1$, the number of cuts of weight at most 
$\alpha c$ is at most $O(n^{2\alpha})$.
\end{theorem}
\noindent While this theorem is extremely useful in bounding the number
of {\em small} cuts in an undirected graph (e.g. in 
sampling~\cite{Karger94a, Karger94b, BenczurK96},
network reliability~\cite{Karger99}, etc.), it does not shed
any light on the distribution of edges according to their 
connectivities in cuts. We generalize the above theorem and show
that though there may be many distinct cuts of a fixed large weight
in a graph, there are a small number of distinct sets of 
edges in these cuts if we restrict our attention to only edges 
with large (standard) connectivity. To state our
theorem precisely, we need to introduce the notion of
{\em $k$-heavy} and {\em $k$-light} edges, and that of the
{\em $k$-projection} of a cut.
\begin{definition}
An edge is said to be $k$-heavy if it has connectivity
at least $k$, and $k$-light otherwise.
The $k$-projection of a cut is the set of $k$-heavy edges
in the cut.
\end{definition}
\noindent 
Since every edge has connectivity at least $c$, 
Theorem~\ref{thm:cut-count} can be interpreted as bounding 
the number of distinct $k$-projections of cuts of size $\alpha k$ 
by $O(n^{2\alpha})$ for $k=c$. We generalize this result to
arbitrary values of $k$.
\begin{theorem}
\label{thm:cut-count-new}
For any undirected graph with edge connectivity $c$ and for any 
$k \geq c$ and any $\alpha \geq 1$, the 
number of distinct $k$-projections of cuts of weight at most 
$\alpha k$ is at most $n^{2\alpha}$.
\end{theorem}
\noindent
We believe this theorem will be of independent interest.

\noindent
\paragraph{Roadmap.} In section~\ref{sec:framework}, we describe
our generic sampling framework, and provide one example of instantiating this
framework that proves Theorem~\ref{thm:algo2} for the unweighted 
case. In section~\ref{sec:proofs}, we prove 
Theorem~\ref{thm:cut-count-new} and use it to prove Theorem~\ref{thm:main},
the main framework theorem stated in section~\ref{sec:framework}.
In section~\ref{sec:polyweights}, 
we give two sampling algorithms for graphs with polynomial 
edge weights: the first algorithm constructs skeletons containing 
$O(\frac{n\log^2 n}{\epsilon^2})$ edges in expectation and
has time complexity $O(m)$, thus proving
Theorem~\ref{thm:algo2} for the polynomial weights case;
the second algorithm constructs skeletons containing 
$O(\frac{n\log n}{\epsilon^2})$ edges in expectation and
has time complexity $O(m\log n)$ for unweighted graphs, and 
$O(m\log^2 n)$ for graphs with polynomial edge weights, thus
proving Theorem \ref{thm:algo2-1}. Combining these two theorems
proves Theorem~\ref{thm:algo}.
In section~\ref{sec:standard}, we prove Theorem~\ref{thm:standard}
and show that results 
on sampling by effective resistances and sampling by strong connectivities 
can also be derived from our framework. Finally, in
section~\ref{sec:weighted}, we give a sampling 
algorithm for graphs with arbitrary edge weights that constructs 
skeletons containing $O(\frac{n\log^2 n}{\epsilon^2})$ edges in 
expectation and has time complexity $O(m\log^2 n)$, thus proving
Theorem~\ref{thm:algo2} for the arbitrary weights case.

\section{The Generic Framework} 
\label{sec:framework}
We describe a generic sampling framework---each of our individual
sampling schemes is obtained by a particular setting of 
parameters of this generic framework. 

Suppose $G = (V, E)$ is an undirected graph where edge $e\in E$ 
has weight $w_e$. We will assume throughout that 
$w_e$ is a positive integer.
Let $G_M = (V, E_M)$ denote the multi-graph
constructed by replacing each edge $e$ by 
$w_e$ unweighted parallel edges $e_1, e_2, \ldots, e_{w_e}$. 
Consider any $\epsilon\in (0, 1]$.
We construct a skeleton $G_{\epsilon}$ where each edge 
$e_{\ell}\in E_M$ is present in graph $G_{\epsilon}$ independently
with probability $p_e$, and if present, it is given a weight of 
$1/p_e$. (For algorithmic efficiency, observe that an identical
skeleton can be created by assigning to edge $e$ 
a weight of $R_e/p_e$ where $R_e$ is generated 
from the binomial distribution $B(w_e,p_e)$; this can be
done in time $O(w_ep_e)$ rather than time $O(w_e)$ (see e.g.~\cite{KachitvichyanukulS88})). 

What values of $p_e$ result in a sparse $G_{\epsilon}$
that satisfies $G_{\epsilon}\in (1\pm \epsilon)G$?
Let $p_e = \min(\frac{96\alpha\ln n}{0.38\lambda_e\epsilon^2}, 1)$, 
where $\alpha$ is independent of $e$ and $\lambda_e$ is some 
parameter of $e$ satisfying $\lambda_e\leq 2^n-1$.  
The exact choice of values for $\alpha$ and the $\lambda_e$'s 
will vary from application to application.
However, we describe below a sufficient condition that characterizes
a good choice of $\alpha$ and $\lambda_e$'s.

To describe this sufficient condition, partition the edges in $G_M$ according to the value of 
$\lambda_e$ into sets $F_0, F_1, \ldots, F_k$ where 
$k = \lfloor\lg \max_{e\in E}\{\lambda_e\}\rfloor\leq n-1$ and
$e_i\in F_j$ iff $2^j\leq \lambda_e\leq 2^{j+1}-1$.
Now, let ${\cal G} = G_0, G_1, G_2, \ldots, G_i = (V, E_i), \ldots, G_k$ 
be a set of subgraphs of $G_M$ (we allow edges of $G_M$ to be replicated
multiple times in the $G_i$s) such that $F_i\subseteq E_i$ for every $i$. 
$\cal G$ is said to be a {\em $(\pi, \alpha)$-certificate} corresponding to
the above choice of $\alpha$ and $\lambda_e$'s if the following properties are satisfied:
\begin{description}
\item[$\pi$-connectivity] 
For $i\geq 0$, any edge $e_{\ell}\in F_i$ is $\pi$-heavy in $G_i$.
\item[$\alpha$-overlap] For any cut $C$ containing $c$ edges in $G_M$, 
let $e^{(C)}_i$ be the number
of edges that cross $C$ in $G_i$. Then, for all cuts $C$,
$\sum_{i=0}^k \frac{e^{(C)}_i 2^{i-1}}{\pi}\leq \alpha c$.
\end{description}
Theorem \ref{thm:main} describes the sufficient condition; its proof
appears later in section~\ref{sec:proofs}. The intuition
for this proof is as follows. Consider all cuts $C$ in $G_M$; restrict each cut to just the
edges in $F_i$ (we do this because edges in $F_i$ have roughly the same sampling probabilities,
which enables an easy application of Chernoff bounds). How many such distinct $F_i$-restricted
cuts are there? Organize all cuts $C$ in $G_M$ into doubling categories, each comprising
cuts with roughly equal values of $e^{(C)}_i$; now using
Theorem \ref{thm:cut-count-new} as applied to $G_i$ and the $\pi$-connectivity property above,
we can conclude that this count is $n^{O(e^{(C)}_i/{\pi})}$ per category.
Next, for a particular cut $C$ and its $F_i$-restriction, we need to apply 
an appropriate Chernoff
bound with a carefully chosen deviation-from-expectation parameter so that
this deviation has  probability at most $n^{-\Omega(e^{(C)}_i/ \pi)}$;
this probability offsets the above count, thereby allowing us to claim that this deviation
holds for all cuts in one doubling category (and the number of categories is not too many,
so the same fact extends across categories as well).
The actual value of this deviation comes out to be $O(\epsilon)\cdot\frac{e^{(C)}_i}{\pi}\cdot\frac{2^{i-1}}{\alpha}$.
The $\alpha$-overlap property now allows us to bound the sum of this deviation over all $i$, $0\leq i\leq k$,
by $\epsilon c$, as required.
 
\begin{theorem}
\label{thm:main}
If there exists a $(\pi, \alpha)$-certificate for a particular choice of $\alpha$ and $\lambda_e$'s , then 
the skeleton $G_{\epsilon}\in (1\pm\epsilon)G$ with probability at least $1 - 4/n$.
Further $G_{\epsilon}$ has $O(\frac{\alpha \log n}{\epsilon^2}\sum_{e\in E} \frac{w_e}{\lambda_e})$
edges in expectation.
\end{theorem}
\noindent

\junk{
We also need to show that $G_{\epsilon}$ is sparse. The following
theorem implies that bounding $\sum_{e} \frac{w_e}{\lambda_e}$ is
sufficient. 
\begin{theorem}
\label{thm:main-sparse}
If $\sum_{e\in E} \frac{w_e}{\lambda_e} = O(S)$ for some $S$, then
the expected number of edges in the skeleton $G_{\epsilon}$ constructed
by the above algorithm is $O(\frac{\alpha S\log n}{\epsilon^2})$.
\end{theorem}
\begin{proof}
The expected number of distinct edges in $G_{\epsilon}$ is
$\sum_{e\in E} 1 - (1 - p_e)^{w_e}\leq \sum_{e} w_e p_e$. The 
theorem follows by substituting the value of $p_e$.
\end{proof}
}

\subsection{A Simple Algorithm for Unweighted Graphs}

We show how we can instantiate the above framework with
specific values of $\alpha$, $\lambda_e$'s 
to obtain a very simple sampling algorithm that runs in $O(m)$ time
and obtains a skeleton of size $O(\frac{n\log^2 n }{\epsilon^2})$.
This proves  Theorem \ref{thm:algo2} for the unweighted case.

In order to present our sampling algorithm, we need to define 
the notion of {\em spanning forests}. As earlier, $G$ denotes a graph with 
integer edge weights $w_e$ for edge $e$ and $G_M$ is the 
unweighted multi-graph where $e$ is replaced with $w_e$ parallel
unweighted edges.
\begin{definition}
A {\em spanning forest} $T$ of $G_M$ (or equivalently of $G$) 
is an (unweighted) acyclic subgraph of $G$ satisfying the property 
that any two vertices are connected in $T$ if and only if they are connected in $G$.
\end{definition}
We partition the set of edges in $G_M$ into a set of 
forests $T_1, T_2, \ldots$ using the following rule: 
{\em $T_i$ is a spanning forest of the graph formed by 
removing all edges in $T_1, T_2, \ldots, T_{i-1}$ from $G_M$
such that for any edge $e\in G$, all its copies in $G_M$
appear in a set of contiguous forests 
$T_{i_e}, T_{i_e+1},\ldots, T_{i_e + w_e-1}$}.
This partitioning technique was introduced by Nagamochi and Ibaraki 
in~\cite{NagamochiI92a}, and these forests are known as 
{\em Nagamochi-Ibaraki forests} (or NI forests). The following is
a basic property of NI forests.
\begin{lemma}[Nagamochi-Ibaraki~\cite{NagamochiI92a, NagamochiI92b}]
\label{lma:ni-con}
For any pair of vertices $u, v$, they are connected in 
NI forests $T_1, T_2, \ldots, T_{k(u, v)}$ for some 
$k(u, v)$ and not connected in any forest $T_j$, for
$j > k(u, v)$.
\end{lemma}
\noindent
Nagamochi and Ibaraki also gave an algorithm for constructing 
NI forests that runs in $O(m+n)$ time if $G_M $ is a simple
graph (i.e. $G$ is unweighted) and 
$O(m+n\log n)$ time otherwise~\cite{NagamochiI92a, NagamochiI92b}. 
Note that our sampling schemes
are relevant only when $m > n\log n$; therefore, the NI forests
can be constructed in $O(m)$ time for all relevant input graphs.

We set 
$\lambda_e$ to the index of the NI forest that $e$ appears in,
and set $\alpha = 2$ and $\pi = 2^{i-1}$. 
For any $i > 0$, let $G_i$ contain all 
edges in NI forests 
$T_{2^{i-1}}, T_{2^{i-1}+1}, \ldots, T_{2^{i+1}-1}$; let
$G_0 = F_0 = T_1$. 
Each edge in $F_i$ appears exactly once in $G_i$, once in $G_{i+1}$,
and does not appear at all in any of the other $G_j$'s, $j\not=i,i+1$.
This proves $\alpha$-overlap. Further, for any edge
$e\in F_i$, $i > 0$, 
Lemma~\ref{lma:ni-con} ensures that the endpoints
of $e$ are connected in each of 
$T_{2^{i-1}}, T_{2^{i-1}+1}, \ldots, T_{2^i-1}$. It follows that
$e$ is $2^{i-1}$-heavy in $G_i$, thereby proving 
$\pi$-connectivity. We can now invoke Theorem \ref{thm:main}
and conclude that this sampling scheme results in 
$G_{\epsilon}\in (1\pm\epsilon)G$ with probability at least $1 - 4/n$.
It remains to bound the number of edges in $G_{\epsilon}$, as follows.

Since $w_e = 1$ for each edge $e$ and the
total number of NI forests $K$ is at most $n^2$, we have
\begin{equation*}
\sum_{e\in E} \frac{w_e}{\lambda_e} = \sum_{e\in E} \frac{1}{\lambda_e} =
\sum_{j=1}^K \sum_{e\in T_j} \frac{1}{\lambda_e} = 
\sum_{j=1}^K \sum_{e\in T_j} \frac{1}{j} \leq
(n-1)\sum_{j=1}^K \frac{1}{j} = O(n\log K) = O(n\log n).
\end{equation*}
It follows from Theorem \ref{thm:main} that 
$G_{\epsilon}$ has  $O(\frac{n\log^2 n }{\epsilon^2})$ edges.

The time complexity for constructing the NI forests is $O(m)$
and that for sampling is $O(1)$ per edge giving another $O(m)$;
so overall, the algorithm takes $O(m)$ time.

\section{Proofs of Main Theorems}
\label{sec:proofs}

In this section, we will first prove Theorem~\ref{thm:cut-count-new}, 
and then use it to prove Theorem~\ref{thm:main}. 
Let us start by defining $k$-heavy and $k$-light vertices.
\begin{definition}
A vertex in an undirected graph is said to be $k$-heavy if at least
one edge incident on the vertex is $k$-heavy; otherwise, the vertex 
is said to be $k$-light.
\end{definition}
\noindent We need the following property of $k$-heavy vertices.
\begin{lemma}
\label{lma:heavy}
The sum of weights of edges incident on a $k$-heavy vertex is at 
least $k$.
\end{lemma}
\begin{proof}
For any $k$-heavy vertex $v$, there exists some other vertex $u$ such that
the maxflow between $u$ and $v$ is at least $k$. Thus, any cut
separating $u$ and $v$ must have weight at least $k$; in particular,
this holds for the cut containing only $v$ on one side.
\end{proof}
\noindent
Suppose $G$ is an any weighted undirected graph. We scale up the weights
of all edges in $G$ uniformly until the weight of every edge is an even
integer; call this graph $G_s$. We replace each edge $e = (u, v)$ of weight
$w_e$ in $G_S$ with $w_e$ parallel unweighted edges between $u$ and $v$ 
to form an unweighted multi-graph $G_M$. Clearly, any cut in $G_M$
has an even number of edges. 
Theorem~\ref{thm:cut-count-new} holds for any value of
$k$ in $G$ if and only if it holds for any even integer $k$ in $G_M$. Therefore,
it suffices to prove Theorem~\ref{thm:cut-count-new} for all even integers $k$ 
on unweighted multigraphs where the weight of every cut
is even. We also assume that $G_M$ is connected; 
if not, the theorem holds for the entire graph since it holds for each 
connected component.

We introduce two operations on undirected
multigraphs: {\em spitting-off} and {\em edge contraction}.
The splitting-off operation was introduced
by Lov\'asz in~\cite{Lovasz74, Lovasz93} (ex. 6.53): 
\begin{definition}
A pair of edges $(s, u)$ and $(u, t)$ are said to be {\em split-off}
in an undirected multigraph if they are replaced by a single edge $(s, t)$.
\end{definition}
\noindent Various properties of the splitting-off operation have been
explored~\cite{Mader78, Mader82, Frank92, Szigeti08}. We need the
following property.
\begin{definition}
For any $k > 0$, a splitting-off operation is said to be 
{\em $k$-preserving} if all edges in the graph 
(except those being split-off) that 
were $k$-heavy before the splitting-off continue to be 
$k$-heavy after the splitting-off.
\end{definition}
\noindent
The following lemma is a corollary of a deep result of 
Mader~\cite{Mader78} for splitting-off edges while maintaining
the maxflows of pairs of vertices; however, we give a much simpler
direct proof of this lemma here.
\begin{lemma}
\label{lma:split-off}
Suppose $G_M$ is an undirected multigraph where every cut contains an even
number of edges. Let $k > 0$ be any even integer. Then, for
any $k$-light non-isolated vertex $u$ in $G_M$, 
there exists a pair of edges $(s, u)$ and $(u, t)$ such that 
splitting-off this pair is $k$-preserving.
\end{lemma}
\begin{proof}
We will prove that for every edge $(s, u)$, there exists an edge $(u, t)$
such that splitting-off this pair of edges retains the following property:
{\em any pair of vertices $x, y$ that were $k$-connected (i.e. had a maxflow
of at least $k$)
before the splitting-off continue to be so after the splitting-off}. 
We define a {\em $k$-separator} to be 
any cut that separates at least one pair of $k$-connected 
vertices, and call a $k$-separator with 
exactly $k$ edges a {\em tight} cut. Since all cuts
have even number of edges and the weight of a cut can decrease by at
most 2 due to a splitting-off operation, we only need to ensure that 
we do not decrease the number of edges in any tight cut when we split-off a 
pair of edges.

Suppose there exists no edge $(u, t)$ such that splitting-off $(s, u)$
and $(u, t)$ retains the $k$-heavy property for all $k$-heavy edges. 
Then, for every 
neighbor $t$ (other than $s$) of $u$, there exists at least one tight
cut having $s, t$ on one side and $u$ on the other. Consider a 
minimum-sized collection of tight cuts $X_1, X_2, \ldots, X_{\ell}$,
where $X_i$ is the subset of vertices on the side of the cut not
containing $u$. If $\ell = 1$, moving $u$ to the side of $X_1$ 
produces a $k$-separator containing less 
than $k$ edges, which is a contradiction. Thus $\ell\geq 2$.
Now,let
\begin{equation*}
A = X_1 \cap X_2; B = X_1 \setminus X_2; C = X_2 \setminus X_1; D = V\setminus (X_1 \cup X_2).
\end{equation*}
Then, $s\in A$ and $u\in D$. Since $X_1$ and $X_2$ are $k$-separators,
either (1) $A$ and $D$ are $k$-separators, or (2) $B$ and $C$ are $k$
separators. In either case, this pair of $k$-separators must be
tight cuts since they contain at least $k$ edges each being $k$-separators
and at most $k$ edges each because their total number of edges is at most
that of $X_1$ and $X_2$. If $A$ and $D$ are tight cuts, 
we can replace cuts $X_1$ and $X_2$
by $D$ in the collection of tight cuts, contradicting minimality of this
collection. On the other hand, if $B$ and $C$ are tight cuts,
the counting argument also shows that there is no edge between $A$
and $D$, contradicting the existence of edge $(s, u)$.
\end{proof}
\noindent
Let us now extend the notion of splitting-off to vertices.
\begin{definition}
A vertex with even degree in an undirected graph is said to
be {\em split-off} if a pair of edges incident on it is
repeatedly split-off until the vertex becomes isolated.
Splitting-off of a vertex is said to be {\em $k$-preserving} 
if each constituent edge splitting-off is $k$-preserving.
\end{definition}
\noindent Note that the number of edges in a cut either stays unchanged 
or decreases by 2 after a splitting-off operation. Thus, if every cut in 
the graph had an even number of edges to start with, then each cut 
continues to have
an even number of edges after a sequence of splitting-off operations. 
Therefore, the following lemma is obtained by repeatedly applying 
Lemma~\ref{lma:split-off} to a $k$-light vertex.
\begin{lemma}
\label{lma:split-off-vertex}
Suppose $G_M$ is an undirected multigraph where the number of edges
in every cut is even. Let $k$ be an even integer. Then, there exists
a $k$-preserving splitting-off of any non-isolated $k$-light vertex 
$u$ in $G_M$.
\end{lemma}
\noindent
Our second operation is {\em edge contraction}.
\begin{definition}
{\em Contraction} of edge $e =(u, v)$ in an undirected multigraph $G$
is defined as merging $u$ and $v$ into a single vertex
(i.e. all edges incident on either $u$ or $v$ are now incident 
on the new vertex instead). 
Any self-loops produced by edges between $u$ and $v$ are discarded.
\end{definition}
\noindent We will now prove Theorem~\ref{thm:cut-count-new}.
\begin{proof}[Proof of Theorem~\ref{thm:cut-count-new}]
We run the following randomized algorithm on multigraph $G_M$:
\begin{enumerate}
\item\label{init} Split-off all $k$-light vertices ensuring the
$k$-preserving property (Lemma~\ref{lma:split-off-vertex}).
\item\label{loop} Contract an edge chosen uniformly at random in the resulting
graph. 
\item\label{split-off} If the contraction produces a $k$-light vertex, split 
it off.\footnote{If an edge between $u$ and $v$ is contracted in step~\ref{loop}, 
all edges that were previously $k$-heavy continue to be so after the 
contraction, except the edges between $u$ and $v$. So, at most one vertex
(the new vertex) becomes $k$-light as a result of this contraction.}
\item If $\leq 2\alpha$ vertices are left, output a random cut; otherwise, go to 
step~\ref{loop}.
\end{enumerate}
\noindent
Consider a cut $C$ that has at most $\alpha k$ edges; let its 
$k$-projection be $S$. In any of the splitting-off operations, no edge 
in $S$ can be split-off since these edges continue to be $k$-heavy 
throughout the execution of the algorithm. So, if no edge crossing 
cut $C$ (either an edge in $G_M$ or one produced by the splitting-off
operations) is
contracted during the execution of the algorithm, then all edges in
$S$ survive till the end. To estimate the probability that no edge
crossing cut $C$ is contracted, let $h_j$ be the number of vertices
left at the beginning of the $j$th iteration. Thus, $h_1$ is the number 
of $k$-heavy vertices in $G_M$ (note that all $k$-light vertices
are split-off initially), and $h_{j+1}$ is 
either $h_j-1$ or $h_j-2$ depending on whether a vertex was split-off 
in step~\ref{split-off} of iteration $j$. Observe that the number of 
edges crossing $C$ cannot increase due to the splitting-off operations.
Further, Lemma~\ref{lma:heavy} asserts that at the beginning of 
iteration $j$, there are at least $h_jk/2$ edges in the graph. 
Thus, the probability that no edge in $C$ is selected for random 
contraction in step~\ref{loop} of iteration $j$ is at least 
$1 - \frac{\alpha k}{h_jk/2} = 1 - \frac{2\alpha}{h_j}$. Then, 
the probability that no edge crossing $C$ is contracted
in the entire execution of the algorithm is at least
\begin{equation*}
\prod_j \left(1 - \frac{2\alpha}{h_j}\right)
\geq \prod_{i=n}^{2\alpha+1} \left(1 - \frac{2\alpha}{i}\right)
= {n\choose 2\alpha}^{-1}.
\end{equation*}
Since there are $2^{2\alpha -1}$ cuts in a graph with $2\alpha$
vertices, the probability that the random cut output by the algorithm 
contains only edges crossing cut $C$ (and therefore $S$ is exactly
the set of $k$-heavy edges in $G_M$ output by the algorithm) is at least
${n\choose 2\alpha}^{-1}2^{1-2\alpha} \geq n^{-2\alpha}$. This is 
true for every distinct $k$-projection of cuts having at most 
$\alpha k$ edges; hence, the total number of such $k$-projections
is at most $n^{2\alpha}$.
\end{proof}

In addition to the above theorem, we need the following 
non-uniform version of Chernoff bounds 
(for Chernoff bounds, see e.g.~\cite{MotwaniR97}) to 
prove Theorem~\ref{thm:main}. (A proof of this theorem
is given in the appendix.)
\begin{theorem}
\label{thm:compression}
Consider any subset $C$ of unweighted edges, where 
each edge $e\in C$ is sampled independently with probability $p_e$
for some $p_e\in [0, 1]$ and given weight $1/p_e$ if selected
in the sample. 
Let the random variable $X_e$ denote the weight of edge $e$ in the
sample; if $e$ is not selected in the sample, then $X_e = 0$. 
Then, for any $p$ such that $p \leq p_e$ for all edges $e$, 
any $\epsilon\in (0, 1]$, and any $N\geq |C|$, the following
bound holds:\footnote{For any event $\cal E$, $\pr[{\cal E}]$
represents the probability of event $\cal E$.}
\begin{equation*}
\pr\left[|\sum_i X_e - |C|| > \epsilon N\right] < 2 e^{-0.38 \epsilon^2 pN}.\qedhere
\end{equation*}
\end{theorem}
\noindent
We will now use Theorem~\ref{thm:cut-count-new} to prove 
Theorem~\ref{thm:main}. (We re-use the notation defined in
section~\ref{sec:framework}.) For any cut $C$ in $G_M$, let 
$F^{(C)}_i = F_i\cap C$ and $E^{(C)}_i = E_i\cap C$
for $0\leq i\leq k$;\footnote{For any cut $C$ and any
set of edges $Z$, $Z\cap C$ denotes the set of edges in
$Z$ that cross cut $C$.} let $f^{(C)}_i = |F^{(C)}_i|$ and
$e^{(C)}_i = |E^{(C)}_i|$. Also, let 
$\widehat{f^{(C)}_i}$ be the expected weight of all edges
in $F^{(C)}_i$ in the skeleton graph 
$G_{\epsilon}$. We first prove a key lemma.
\begin{lemma}
\label{lma:main}
For any fixed $i$, with probability at least $1 - \frac{4}{n^2}$,
\begin{equation*}
|f^{(C)}_i-\widehat{f^{(C)}_i}| \leq \frac{\epsilon}{2} \max\left(\frac{e^{(C)}_i 2^{i-1}}{\pi \alpha}, f^{(C)}_i\right)
\end{equation*}
for all cuts $C$ in $G_M$.
\end{lemma}
\begin{proof}
By the $\pi$-connectivity property, any edge $e\in F_i$ is 
$\pi$-heavy in $G_i$ for any $i\geq 0$. Therefore, $e^{(C)}_i\geq \pi$.
Let ${\cal C}_{ij}$ be the set of all cuts $C$ such that 
$\pi 2^j\leq e^{(C)}_i\leq \pi 2^{j+1}-1$, $j\geq 0$. 
We will prove that with probability at least $1 - 2n^{-2^{j+1}}$,
all cuts in ${\cal C}_{ij}$ satisfy the property of the lemma.
Then, the lemma follows by using the union bound over $j$ 
(keeping $i$ fixed) since 
$2n^{-2} + 2n^{-4} + \ldots + 2n^{-2j} + \ldots \leq 4n^{-2}$.

We now prove the above claim for cuts $C\in {\cal C}_{ij}$. 
Let $X^{(C)}_i$ denote the set of edges in $F^{(C)}_i$
that are sampled with probability strictly less than 1;
correspondingly, let $x^{(C)}_i = |X^{(C)}_i|$ and let 
$\widehat{x^{(C)}_i}$ be the total weight of edges in 
$X^{(C)}_i$ in the skeleton graph $G_{\epsilon}$. Since edges in
$F^{(C)}_i \setminus X^{(C)}_i$ have a weight of exactly 1 in 
$G_{\epsilon}$, it is sufficient to show that with probability
at least $1 - 2n^{-2^{j+1}}$,
$|x^{(C)}_i-\widehat{x^{(C)}_i}| \leq \left(\frac{\epsilon}{2}\right) \max\left(\frac{e^{(C)}_i 2^{i-1}}{\pi \alpha}, x^{(C)}_i\right)$
for all cuts $C\in {\cal C}_{ij}$. Since each edge $e\in X^{(C)}_i$
has $\lambda_e < 2^{i+1}$, we can use Theorem~\ref{thm:compression}
with the lower bound on probabilities 
$p = \frac{96\alpha\ln n}{0.38 \cdot 2^{i+1}\epsilon^2}$. 
There are two cases. In the first case, suppose 
$x^{(C)}_i\leq \frac{e^{(C)}_i 2^{i-1}}{\pi \alpha}$. Then, for any
$X^{(C)}_i$ where $C\in {\cal C}_{ij}$, by Theorem~\ref{thm:compression}, 
we have
\begin{equation*}
\pr\left[\left|x^{(C)}_i-\widehat{x^{(C)}_i}\right| > \left(\frac{\epsilon}{2}\right)\frac{e^{(C)}_i 2^{i-1}}{\pi \alpha}\right] <
2e^{-0.38\frac{\epsilon^2}{4} \left(\frac{96\alpha\ln n}{0.38\cdot 2^{i+1}\epsilon^2}\right) \frac{e^{(C)}_i 2^{i-1}}{\pi \alpha}} \leq
2e^{-\frac{6 e^{(C)}_i\ln n}{\pi}} \leq
2e^{-6\cdot 2^j\ln n},
\end{equation*}
since $e^{(C)}_i\geq \pi 2^j$ for any $C\in {\cal C}_{ij}$.
In the second case, suppose
$x^{(C)}_i > \frac{e^{(C)}_i 2^{i-1}}{\pi \alpha}$. Then, for any
$X^{(C)}_i$ where $C\in {\cal C}_{ij}$, by Theorem~\ref{thm:compression}, 
we have
\begin{equation*}
\pr\left[\left|x^{(C)}_i-\widehat{x^{(C)}_i}\right| > \left(\frac{\epsilon}{2}\right) x^{(C)}_i\right] <
2e^{-0.38\frac{\epsilon^2}{4} \left(\frac{96\alpha\ln n}{0.38\cdot 2^{i+1}\epsilon^2}\right) x^{(C)}_i} <
2e^{-\frac{6 e^{(C)}_i\ln n}{\pi}} \leq
2e^{-6\cdot 2^j\ln n},
\end{equation*}
since $x^{(C)}_i > \frac{e^{(C)}_i 2^{i-1}}{\pi \alpha}\geq \frac{2^{i+j-1}}{\alpha}$ 
for any $C\in {\cal C}_{ij}$. Thus, we have proved that
\begin{equation*}
\pr\left[\left|x^{(C)}_i-\widehat{x^{(C)}_i}\right| > \left(\frac{\epsilon}{2}\right) \max\left(\frac{e^{(C)}_i 2^{i-1}}{\pi \alpha}, x^{(C)}_i\right)\right] < 
2e^{-6\cdot 2^j\ln n} = 
2n^{-6\cdot 2^j}
\end{equation*}
for any cut $C\in {\cal C}_{ij}$. Now, by the $\pi$-connectivity 
property, we know that edges in $F^{(C)}_i$, and therefore those in 
$X^{(C)}_i$, are $\pi$-heavy in $G_i$. Therefore, by 
Theorem~\ref{thm:cut-count-new}, the number of distinct $X^{(C)}_i$
sets for cuts $C\in {\cal C}_{ij}$ is at most 
$n^{2\left(\frac{\pi 2^{j+1}}{\pi}\right)} = n^{4\cdot 2^j}$.
Using the union bound over these distinct $X^{(C)}_i$ edge sets,
we conclude that with probability at least $1 - 2n^{-2^{j+1}}$,
all cuts in ${\cal C}_{ij}$ satisfy the property of the lemma.
\end{proof}
\noindent
We now use the above lemma to prove Theorem~\ref{thm:main}.
\begin{proof}[Proof of Theorem~\ref{thm:main}]
For any cut $C$ in $G_M$, let $c$ be the number of 
edges in $C$; correspondingly, let $\hat{c}$ be the total weight of the
edges crossing cut $C$ in the skeleton graph $G_{\epsilon}$. 
Since $k\leq n-1$, we apply the union bound to the property
from Lemma~\ref{lma:main} over the different
values of $i$ to conclude that with probability at least
$1-\frac{4}{n}$, we have 
$\sum_{i=0}^k |\widehat{f^{(C)}_i} - f^{(C)}_i| \leq 
\sum_{i=0}^k \left(\frac{\epsilon}{2}\right)\max\left(\frac{e^{(C)}_i 2^{i-1}}{\pi \alpha}, f^{(C)}_i\right)$ 
for all cuts $C$ in $G_M$. Then, with probability at least $1-\frac{4}{n}$,
\begin{equation*}
|\hat{c} - c| =
|\sum_{i=0}^k \widehat{f^{(C)}_i} - \sum_{i=0}^k f^{(C)}_i| \leq
\sum_{i=0}^k |\widehat{f^{(C)}_i} - f^{(C)}_i| \leq
\frac{\epsilon}{2}\sum_{i=0}^k \max\left(\frac{e^{(C)}_i 2^{i-1}}{\pi \alpha}, f^{(C)}_i\right) \leq
\frac{\epsilon}{2}\left(\sum_{i=0}^k  \frac{e^{(C)}_i 2^{i-1}}{\pi \alpha} + \sum_{i=0}^k f^{(C)}_i\right) \leq
\epsilon c, 
\end{equation*}
since $\sum_{i=0}^k  \frac{e^{(C)}_i 2^{i-1}}{\pi \alpha}\leq c$ by the 
$\alpha$-overlap property and $\sum_{i=0}^k f^{(C)}_i\leq c$ since $F^{(C)}_i$'s
form a partition of the edges in $C$.

We now prove the size bound on $G_{\epsilon}$.
The expected number of distinct edges in $G_{\epsilon}$ is
\begin{equation*}
\sum_{e\in E} 1 - (1 - p_e)^{w_e}\leq \sum_{e} w_e p_e. 
\end{equation*}
The bound follows by substituting the value of $p_e$.
\end{proof}

\section{Sampling in Graphs with Polynomial Edge Weights}
\label{sec:polyweights}

In this section, we will give an algorithm for 
sampling in undirected weighted graphs, where the
weight of every edge is an integer bounded by $n^d$ 
for a fixed constant $d > 0$. The algorithm constructs
a skeleton graph containing $O(\frac{n\log n}{\epsilon^2})$
edges in expectation and has time complexity 
$O(m + \frac{n\log^4 n}{\epsilon^2})$. 
Our strategy, as outlined in the introduction,
has two steps: first we run an algorithm that
constructs a skeleton graph with 
$O(\frac{n\log^2 n}{\epsilon^2})$ edges in expectation
and has time complexity $O(m)$; 
then, we run a different algorithm that constructs 
a sparser skeleton containing $O(\frac{n\log n}{\epsilon^2})$ 
edges in expectation on the skeleton graph constructed in the
first step. The second algorithm takes time $O(m\log^2 n)$
on a graph with $m$ edges and therefore 
$O(\frac{n\log^4 n}{\epsilon^2})$ time on the skeleton
graph produced in the first step. To ensure that the final
skeleton graph is in $(1\pm \epsilon)G$, we choose $\epsilon/3$
as the error parameter for each algorithm. As an additional
observation, we show that the time complexity of the 
second algorithm improves
to $O(m\log n)$ if its input graph is unweighted.

We will describe both these algorithms for an input graph
$G$, where the weight $w_e$ of every edge $e$ is an integer
bounded by $n^d$ for a fixed constant $d > 0$. Note that
the input graph to the second algorithm in the above two-step 
sampling scheme may have fractional weights. However, we can
scale up all weights uniformly until they are integral, and the scaled
weights continue to be bounded by some fixed polynomial in 
$n$. Once the skeleton graph is obtained, we scale all
weights down uniformly to obtain the final skeleton graph. 
The unweighted multigraph constructed by replacing each edge $e$ with 
$w_e$ parallel unweighted edges 
$e_i, e_2, \ldots, e_{w_e}$ between $u$ and $v$ is
denoted by $G_M$. Also, $T_1, T_2, \ldots$ denotes a
set of NI forests of $G_M$; edge $e_j$ appears in
forest $T_{i_e+j-1}$, where $1\leq j\leq w_e$. Thus,
the copies of edge $e$ appear in NI forests 
$T_{i_e}, T_{i_e+1}, \ldots, T_{i_e+w_e-1}$. For both
algorithms, we will use the generic sampling scheme
described in section~\ref{sec:framework}.

\noindent
\paragraph{Algorithm for Step 1.}
For any edge $e = (u, v)$, we choose $\lambda_e = i_e + w_e -1$,
i.e. the index of the last NI forest where a copy of $e$ appears; 
also set $\alpha = 2$ and $\pi = 2^{i-1}$. 
For any $i\geq 1$, define $G_i$ to be the graph containing 
all edges in NI forests 
$T_{2^{i-1}}, T_{2^{i-1}+1}, \ldots, T_{2^i-1}$ (call this set
of edges $Y_i$) and all edges in $F_i$, i.e. all edges $e$ with 
$2^i\leq \lambda_e\leq 2^{i+1}-1$. Let $G_0$ only contain edges 
in $F_0$. For any $i\not= j$,
$F_i\cap F_j = Y_i\cap Y_j = \emptyset$; thus,
each edge appears in $G_i$ for at most two different values of 
$i$, proving $\alpha$-overlap. Further, for any edge
$e\in F_i$, Lemma~\ref{lma:ni-con} ensures that the endpoints
of $e$ are connected in each of 
$T_{2^{i-1}}, T_{2^{i-1}+1}, \ldots, T_{2^i-1}$. It follows that
$e$ is $2^{i-1}$-heavy in $G_i$, thereby proving 
$\pi$-connectivity. 

We now prove the size bound. For any edge $e'\in E_M$, let
$t(e')$ be the index of the NI forest it appears in.
Then, 
\begin{equation*}
\sum_{e\in E} \frac{w_e}{\lambda_e} =
\sum_{e\in E} \sum_{j=1}^{w_e}\frac{1}{i_e+w_e-1}\leq
\sum_{e\in E} \sum_{j=1}^{w_e}\frac{1}{i_e+j-1} =
\sum_{e'\in E_M} \frac{1}{t(e')} =
\sum_{\ell=1}^K \sum_{e'\in T_{\ell}} \frac{1}{\ell} \leq
(n-1)\sum_{\ell=1}^K \frac{1}{\ell} = O(n\log K) = O(n\log n),
\end{equation*}
where the last step follows from the observation that
the total number of NI forests $K$ is at most $n^{d+2}$,
where $d$ is a constant.
Using Theorem~\ref{thm:main}, we conclude that the skeleton
graph $G_{\epsilon}$ constructed by the above algorithm has 
$O(\frac{n\log^2 n}{\epsilon^2})$ edges in expectation and
is in $(1\pm \epsilon)G$ whp. 

\noindent\paragraph{Time Complexity.}
The time complexity for constructing the NI forests, and therefore
figuring out $p_e$ values is $O(m + n\log n)$.
We sample each edge $e$ by setting its weight in the skeleton
$G_{\epsilon}$ to $r_e/p_e$, where $r_e$ is drawn 
randomly from the Binomial distribution with parameters 
$w_e$ and $p_e$. This is clearly equivalent to the sampling
scheme described above, and can be done in $w_e p_e$ expected
time for each edge $e$ (see e.g.~\cite{KachitvichyanukulS88}), 
and therefore
$O(\frac{n\log^2 n}{\epsilon^2})$ time overall. 
Since $m > \frac{n\log^2 n}{\epsilon^2}$ for this algorithm
to be invoked, the overall time complexity of the 
algorithm is $O(m)$. 

\noindent
\paragraph{Algorithm for Step 2.}
Before describing our second sampling algorithm, we define
the following operation on graphs. (Recall the definition
of edge contraction given in section~\ref{sec:proofs}.)
\begin{definition}
Let $G = (V, E)$ be an undirected graph, and let 
$V_1, V_2, \ldots, V_k$ be a partition of the vertices
in $G$ such that for each $V_i$, the induced graph of 
$G$ on $V_i$ is connected. Then, {\em shrinking} $G$ 
with respect to $V_1, V_2, \ldots, V_k$
produces the graph formed by contracting all edges
between vertices in the same $V_i$ for all $i$.
\end{definition}
\noindent
Our sampling algorithm uses our generic sampling scheme
where $\lambda_e$ is determined using the following 
algorithm. Here $H_c = (V_c, E_c)$ is a 
graph variable representing a weighted graph. 
The algorithm is described recursively; we call
\verb|SetLambda|($G, 0$) to execute it.\\

\noindent
\verb|SetLambda|($H, i$)
\begin{enumerate}
\item Set $H_c = H$
\item\label{inner} If total weight of edges in $E_c$ is at most $|V_c|\cdot 2^{i+1}$, then
	\begin{enumerate}
	\item Set $\lambda_e = 2^i$ for all edges $e\in E_c$
	\item Remove all edges in $E_c$ from $H$; suppose $H$ splits into connected components $H_1, H_2, \ldots, H_k$
	\item	For each $H_j$ containing at least 2 vertices, call \verb|SetLambda|($H_j, i+1$)
	\end{enumerate}
	Else,
	\begin{enumerate}
	\item Construct $2^i + 1$ NI forests $T_1, T_2, \ldots, T_{2^i+1}$ for $H_c$
	\item Shrink $H_c$ wrt the connected components in $T_{2^i+1}$; update $V_c$ and $E_c$ accordingly
	\item Go to step~\ref{inner}
	\end{enumerate}
\end{enumerate}
Also, set $\alpha = 4$ and $\pi = 2^k$ where 
$k = \lfloor \lg \max_{e\in E}\{\lambda_e\}\rfloor$. 
For any $r$, 
recall that $F_r$ contains all $w_e$ unweighted copies of
edge $e$ from $G_M$, where $e$ satisfies 
$2^r\leq \lambda_e\leq 2^{r+1}-1$. For any $i\geq 1$, let
$G_i$ contain all edges in $F_r$ for all $r\geq i-1$, where
each edge in $F_r$ is replicated $2^{k-r+1}$ times in $G_i$; 
let $G_0$ contain edges of $F_0$ where each edge is 
replicated $2^k$ times. We need
the following lemma to prove that 
$\pi$-connectivity is satisfied.
\begin{lemma}
\label{lma:sparse-con}
For any $j\geq 1$, consider any edge $e\in F_j$, i.e. an
edge $e$ for which the above algorithm sets 
$\lambda_e = 2^j$. Then, $e$ is $2^{j-1}$-heavy
in the graph $\cup_{r\geq j-1} F_r$. 
\end{lemma}
\begin{proof}
For any edge $e$ in $F_j$, let $G_e = (V_e, E_e)$ be the
component of $G$ containing $e$ such that 
\verb|SetLambda|($G_e, j-1$) was executed. We will
show that $e$ is $2^{j-1}$-heavy in $G_e$; since 
$G_e$ is a subgraph of $G$, the lemma follows. In the 
execution of \verb|SetLambda|($G_e, j-1$), there 
are multiple shrinking operations, each of them 
comprising the contracting of a set of edges. We claim 
that any such contracted edge is $2^{j-1}$-heavy
in $G_e$; it follows that any two vertices $u$ and
$v$ that got shrunk into a single vertex are 
$2^{j-1}$-connected in $G_e$. 
 
Let $G_e$ have $k$ shrinking phases; let the graph
produced after shrinking phase $r$ be $G_{e, r}$. 
We now prove that all edges contracted in phase $r$
must be $2^{j-1}$-connected in $G_e$ by induction 
on $r$. For $r=1$, since $e$ appears in the
$(2^{j-1}+1)$st NI forest of phase 1, $e$ is 
$2^{j-1}$-connected in $G_e$. For the inductive
step, assume that the property holds for phases
$1, 2, \ldots, r$. Any edge that is contracted in 
phase $r+1$ appears in the $(2^{j-1}+1)$st NI forest 
of phase $r+1$; therefore, $e$ is $2^{j-1}$-connected 
in $G_{e, r}$. By the inductive hypothesis, 
all edges of $G_e$ contracted 
in previous phases are $2^{j-1}$-heavy in $G_e$; 
therefore, an edge that is $2^{j-1}$-heavy in $G_{e, r}$
must have been $2^{j-1}$-heavy in $G_e$.
\end{proof}
\noindent
Consider any cut $C$ in $G$ containing an edge $e\in F_i$ for any
$i\geq 0$. Let the
corresponding cut (i.e. with the same bipartition of
vertices) in $G_i$ be $C_i$. We need to show that
the number of edges in $C_i$ is at least $2^k$ to 
prove $\pi$-connectivity. If $i = 0$, $e$ is replicated
$2^k$ times in $G_0$ thereby proving the property.
For $i\geq 1$, 
let the maximum $\lambda_a$ of an edge $a$ in $C$
be $k_C$, where $2^j\leq k_C\leq 2^{j+1}-1$ for some
$j\geq i$. By the above lemma, $C_i$ contains at least
$2^{j-1}$ distinct edges of $G$, each of which is replicated
at least $2^{k-j+1}$ times. Thus, $C_i$ contains at least
$2^k$ edges. 

We now prove $\alpha$-overlap. For any cut $C$, recall that
$f^{(C)}_i$ and $e^{(C)}_i$ respectively denote the number
of edges in $F_i\cap C$ and in $C_i$ (where $C_i$ is as
defined in the previous paragraph) respectively. Then, 
\begin{eqnarray*}
\sum_{i=0}^k \frac{e^{(C)}_i 2^{i-1}}{\pi} & = &
\frac{e^{(C)}_0}{2\pi} + \sum_{i=1}^k \frac{e^{(C)}_i 2^{i-1}}{\pi} =
\frac{f^{(C)}_0 2^k}{2^{k+1}} + \sum_{i=1}^k \frac{f^{(C)}_i 2^{k-r+1} 2^{i-1}}{2^k} =
\frac{f^{(C)}_0}{2} + \sum_{i=1}^k \sum_{r=i-1}^k \frac{f^{(C)}_r}{2^{r-i}} \\
& \leq &
f^{(C)}_0 + \sum_{r=0}^k \sum_{i=1}^{r+1} \frac{f^{(C)}_r}{2^{r-i}} \leq 
3 f^{(C)}_0 + \sum_{r=1}^k f^{(C)}_r \sum_{i=1}^{r+1} \frac{1}{2^{r-i}} \leq
4 f^{(C)}_0 + 4\sum_{r=1}^k f^{(C)}_r \leq 4c.
\end{eqnarray*}

Define $D_i$ to be the set of connected components in the graph
$G\setminus (F_0\cup F_1\cup \ldots \cup F_{i-1})$
for any $i\geq 1$; let $D_0$ be the single connected component 
in $G$. For any $i\geq 0$, if any connected component in $D_i$ 
remains intact in $D_{i+1}$, then there is no edge from that 
connected component in $F_i$. On the other hand, if a component
in $D_i$ splits into $\eta$ components in $D_{i+1}$, then the 
algorithm explicitly ensures that the number of edges in $F_i$ 
from that connected component is at most $\eta 2^{i+1}$. Since each
such edge has $\lambda_e = \frac{1}{2^i}$, the contribution
of these edges to the sum $\sum_{e\in E} \frac{w_e}{\lambda_e}$
is at most $2\eta \leq 4(\eta-1)$ (since $\eta\geq 2$). But, 
$\eta-1$ is the increase in the number of components arising from 
this single component. Therefore, if $d_i = |D_i|$, then 
\begin{equation*}
\sum_{e}\frac{w_e}{\lambda_e} \leq \sum_{i=0}^k 4(d_{i+1} - d_i) \leq 4n
\end{equation*}
since ultimately we have $n$ singleton components.
Using Theorem~\ref{thm:main}, we conclude that the skeleton
graph $G_{\epsilon}$ constructed by the above algorithm has 
$O(\frac{n\log n}{\epsilon^2})$ edges in expectation and
is in $(1\pm \epsilon)G$ whp. 

\noindent
\paragraph{Time Complexity.}
We show below that the algorithm 
to find values of $\lambda_e$ can be implemented in $O(m\log n)$ 
time for unweighted graphs, and $O(m\log^2 n)$ time for graphs with 
polynomial edge weights. Once we have obtained the sampling
probabilities, we use the same trick as in the previous
algorithm, i.e. sample from a Binomial distribution,
to produce the skeleton in $O(\frac{n\log n}{\epsilon^2})$ 
additional time. Since the algorithm is invoked only if 
$m > \frac{n\log n}{\epsilon^2}$, the total running time is
$O(m\log n)$ if $G$ is unweighted and $O(m\log^2 n)$ otherwise.

We now determine the time complexity for finding the values of $\lambda_e$.
Consider one call to \verb|SetLambda(H,i)| which begins with $H=(V,E)$ and
let $H_c=(V_c,E_c)$ denote the graph $H$ as it evolves over the various iterations in this procedure.
Each iteration of steps (a) and (b) in the else block takes 
$O(|V_c|\log n+|E_c|)$ time.
We show that the number of vertices halves in each iteration (save the last)
and therefore the total time over all iterations is $O(|V|\log n+|E|\log n)$.
Since we are dealing with the case of polynomial edge weights,
the depth of recursion is $O(\log n)$.
Therefore, over all recursive calls, the time comes to $O(n\log^2 n+ m\log^2 n)=O(m\log^2 n)$.

To see that the number of vertices halves from one iteration to the next, consider
an iteration that begins with $E_c$ having weight at least $|V_c|\cdot 2^{i+1}$.
$E_c$ for the next iteration (denoted by $E'_c$) comprises only edges in the
first $2^i$ NI forests constructed in the current iteration.
So the total weight of edges in $E'_c$ is at most $|V_c|\cdot 2^i$.
If this is not the last iteration, then this weight exceeds $|V'_c|\cdot 2^{i+1}$.
It follows that $|V'_c|\leq |V_c|/2$, as required.

From the above description,  note that for the unweighted case, $|E'_c|\leq |E_c|/2$,
and therefore the time taken over all iterations in one recursive call is $O(|V|+|E|)$.
Over all recursive calls this comes to $O(m\log n)$.

\section{Sampling Schemes using various Connectivity Parameters}
\label{sec:standard}

In this section, we present several sampling schemes
using various measures of connectivity. Some of these
results were previously known; however, we will show
that these results follow as simple corollaries of our
generic sampling scheme whereas the original proofs 
were specific to each scheme and substantially more
complicated. The algorithms for implementing these 
schemes are less efficient than the algorithms that
we have previously presented; therefore we restrict 
ourselves to structural results in this section. As
earlier, $G$ is the weighted input graph (with 
arbitrary integer weights); $G_M$ is the corresponding
unweighted multigraph; $T_1, T_2, \ldots, T_K$ is a set
of NI forests of $G_M$.

\subsection{Sampling using Standard Connectivities}
For any edge $e = (u, v)$, set $\lambda_e$ to the standard
connectivity of the edge; also set $\alpha = 3+\lg n$ and
$\pi = 2^{i-1}$. $F_i$ is defined as the set of all edges $e$ with
$2^i\leq \lambda_e\leq 2^{i+1}-1$ for any $i\geq 0$.
For any $i\geq 1+\lg n$, let $G_i$ contain all edges in NI forests 
$T_{2^{i-1-\lg n}}, T_{2^{i-1-\log n}+1}, \ldots, T_{2^{i+1}-1}$ 
and all edges in $F_i$. For $i\leq \lg n$, $G_i$ contains
all edges in $T_1, T_2, \ldots, T_i$ and all edges in 
$F_i$. For any $i\geq 0$, let $Y_i$ denote the set of edges 
in $G_i$ but not in $F_i$. For any $i\not= j$,
$F_i\cap F_j = \emptyset$ and each edge appears in $Y_i$ for 
at most $2 + \log n$ different values of $i$; this proves 
$\alpha$-overlap. To prove $\pi$-connectivity, we note that
Lemma~\ref{lma:ni-con} ensures that 
for any pair of vertices $u, v$ with maximum flow $f(u, v)$ 
and for any $k\geq 1$, $u, v$ are at least 
$\min(f(u, v), k)$-connected in the union of the first 
$k$ NI forests, i.e. in $T_1\cup T_2\cup\ldots T_k$.
Thus, any edge $e\in F_i$ is at least $2^i$-heavy in the union of
the NI forests $T_1, T_2, \ldots, T_{2^{i+1}-1}$. Since
there are at most $2^{i-1}$ edges overall in 
$T_1, T_2, \ldots, T_{2^{i-1-\lg n}-1}$, any edge $e\in F_i$
is $2^{i-1}$-heavy in $G_i$. This proves $\pi$-connectivity.

We now prove the size bound. The next lemma is similar to its
corresponding lemma for strong connectivity in~\cite{BenczurK96}.
\begin{lemma}
\label{lma:con-sizebound-weighted}
Suppose $G$ is an undirected graph where edge $e$ has weight
$w_e$ and standard connectivity $k_e$. Then, 
$\sum_e \frac{w_e}{k_e}\leq n-1$.
\end{lemma}
\begin{proof}
We use induction on the number of vertices in the graph. 
For a graph with a single vertex and no edge, the lemma
holds vacuously. Now, suppose the lemma holds for all 
graphs with at most $n-1$ vertices. Let $C$ be a minimum 
cut in $G$, and let $\lambda$ be its weight. For any edge
$e\in C$, $k_e = \lambda$. Thus, 
$\sum_{e\in C} \frac{w_e}{k_e} = 1$. We remove all edges
in $C$ from $G$; this splits $G$ into two connected 
components $G_1$ and $G_2$ with $n_1$ and $n_2$ vertices
respectively, where $n_1, n_2\leq n-1$. Further, the
standard connectivity of each edge in $G_1, G_2$ is 
at most that in $G$. Using the inductive hypothesis,
we conclude that 
$\sum_{e\in G_1} \frac{w_e}{k_e}\leq n_1-1$ and 
$\sum_{e\in G_2} \frac{w_e}{k_e}\leq n_2-1$. We conclude
that 
\begin{equation*}
\sum_e \frac{w_e}{k_e}\leq n_1 - 1 + n_2 -1 + 1 = n-1.
\end{equation*}
\end{proof}
\noindent
Using Theorem~\ref{thm:main}, we conclude that the
expected number of edges in the skeleton graph $G_{\epsilon}$
is $O(\frac{n\log^2 n}{\epsilon^2})$ and 
$G_{\epsilon}\in (1\pm \epsilon)G$ whp.

\subsection{Sampling using Effective Resistances}

For any edge $e = (u, v)$, set $\lambda_e$ to the effective
conductance of the edge, i.e. $\lambda_e = \frac{1}{R_e}$ 
where $R_e$ is the effective resistance of edge $e$. The next
two lemmas imply that the skeleton 
$G_{\epsilon}\in (1\pm\epsilon)G$ whp.
\begin{lemma}
Suppose that a sampling scheme (that uses the generic sampling
scheme) has $\lambda_e \leq k_e$ for each edge $e$ in graph
$G$, where $k_e$ is the standard connectivity of $e$ in $G$.
Then, the skeleton constructed is in $(1\pm\epsilon)G$ whp.
\end{lemma}
\begin{proof}
We use the same definition of $\alpha$, $\pi$ and $G_i$s as 
in the sampling scheme with standard connectivities, and verify
that $\pi$-connectivity and $\alpha$-overlap continue to be
satisfied.
\end{proof}
\begin{lemma}
Suppose edge $e$ in an undirected graph $G$ has standard 
connectivity $k_e$ and effective resistance $R_e$. 
Then, $\frac{1}{R_e}\leq k_e$.
\end{lemma}
\begin{proof}
Consider a cut $C$ of weight $k_e$ separating the terminals of edge $e$. We contract
each side of this cut into a single vertex. In other words, we reduce the resistance
on each edge, other than those in $C$, to 0. By 
Rayleigh's monotonicity principle (e.g.~\cite{DoyleS84}), 
the effective resistance of $e$ does not increase due to this transformation. Since the
effective resistance of $e$ after the transformation is $1/k_e$, $R_e\geq 1/k_e$ in
the original graph.
\end{proof}
 
The size bound follows from the following well-known fact (see 
e.g.~\cite{SpielmanS08}).\footnote{There are many proofs of this 
fact, e.g. use linearity of expectation coupled with the fact that effective 
resistance of an edge is the probability that the edge is in a 
random spanning tree of the graph~\cite{Bollobas98}.}
\begin{fact}
If $R_e$ is the effective resistance of edge $e$ with weight $w_e$ in 
an undirected graph, then $\sum_e w_e R_e\leq n-1$.
\end{fact}
\noindent
It follows from Theorem~\ref{thm:main} that the 
expected number of edges in skeleton $G_{\epsilon}$
is $O(\frac{n\log^2 n}{\epsilon^2})$.

\subsection{Sampling using Strong Connectivities}

For any edge $e$, set $\lambda_e$ to the strong
connectivity of the edge; set $\alpha = 1$ and 
$\pi = 2^k$, where 
$k = \lfloor\lg \max_{e\in E}\{\lambda_e\}\rfloor$. Let
$G_i$ contain all edges in $F_r$ for all $r\geq i$, where
each edge in $F_r$ is replicated $2^{k-r}$ times. We use the
following property of strong connectivities that also appears
in~\cite{BenczurK96}. 
\begin{lemma}
In any undirected graph $G$, if an edge $e$ has strong 
connectivity $k$, then $e$ continues to have strong 
connectivity $k$ even after all edges with strong connectivity 
strictly less than $k$ have been removed from $G$.
\end{lemma}
\noindent
Consider any cut $C$ with an edge $e\in F_i$. Let the
corresponding cut (i.e. with the same bi-partition of
vertices) in $G_i$ be $C_i$. We need to show that
the number of edges in $C_i$ is at least $2^k$ to 
prove $\pi$-connectivity. 
Let the maximum strong connectivity of an edge in $C$
be $k_C$, where $2^j\leq k_C\leq 2^{j+1}-1$ for some
$j\geq i$. By the above lemma, $C_i$ contains at least
$2^j$ distinct edges of $G$, each of which is replicated
at least $2^{k-j}$ times. Thus, $C_i$ contains at least
$2^k$ edges. 

We now prove $\alpha$-overlap. For any cut $C$, recall that
$f^{(C)}_i$ and $e^{(C)}_i$ respectively denote the number
of edges in $F_i\cap C$ and in $C_i$ (where $C_i$ is as
defined in the previous paragraph) respectively. Then, 
\begin{equation*}
\sum_{i=0}^k \frac{e^{(C)}_i 2^{i-1}}{\pi} =
\sum_{i=0}^k \sum_{r=i}^k \frac{f^{(C)}_r 2^{k-r} 2^{i-1}}{2^k} =
\sum_{i=0}^k \sum_{r=i}^k \frac{f^{(C)}_r}{2^{r-i+1}} =
\sum_{r=0}^k \sum_{i=0}^r \frac{f^{(C)}_r}{2^{r-i+1}} =
\sum_{r=0}^k f^{(C)}_r \sum_{i=0}^r \frac{1}{2^{r-i+1}} <
\sum_{r=0}^k f^{(C)}_r = c.
\end{equation*}
 
The size bound follows from the following lemma due
to Bencz\'ur and Karger.
\begin{lemma}[Bencz\'ur-Karger~\cite{BenczurK96}]
If $k_e$ is the strong connectivity of edge $e$ with weight 
$w_e$ in an undirected graph, then $\sum_e \frac{w_e}{k_e}\leq n-1$.
\end{lemma}
\noindent
It follows from Theorem~\ref{thm:main} that the 
expected number of edges in the skeleton graph 
$G_{\epsilon}$ is $O(\frac{n\log n}{\epsilon^2})$ and that 
$G_{\epsilon}\in (1\pm\epsilon)G$ whp.

\section{Sampling in Graphs with Arbitrary Edge Weights} 
\label{sec:weighted}

Unfortunately, the algorithms presented earlier for sampling in
a graph with polynomial edge weights fail if the edge weights 
are arbitrary. In particular, we can no longer guarantee that
the expected number of edges in a skeleton graph constructed
by these algorithms is $\tilde{O}(n/\epsilon^2)$, even though 
it continues to approximately preserve the weight of all cuts whp. 
Therefore, we need to modify our techniques to restore the size
bounds, as described below.

We sort the edges in decreasing order of their weight, breaking ties
arbitrarily. 
We add edges to the NI forests in this sorted order, i.e. when edge $e$ is
being added, the NI forests contain all edges of weight greater than $e$.
To insert $e = (u, v)$, we find the NI forest 
with the minimum index where $u$ and $v$ are not connected; call this
index $i_e$. Then, $e$ is inserted in NI forests 
$T_{i_e}, T_{i_e+1}, \ldots, T_{i_e+w_e-1}$. Note that this does not 
produce any cycle in the NI forests since Lemma~\ref{lma:ni-con} ensures
that if $u, v$ are disconnected in $T_{i_e}$, then they are not connected 
in $T_k$ for any $k\geq i_e$. 

For any edge $e = (u, v)$, set $\lambda_e$ to the index of the 
first NI forest where edge $e$ is inserted, i.e. $\lambda_e = i_e$;
also set $\alpha = 2$ and $\pi = 2^{i-1}$. 
For any $i\geq 1$, let $G_i$ contain all edges in NI forests 
$T_{2^{i-1}}, T_{2^{i-1}+1}, \ldots, T_{2^i-1}$ (call this set
of edges $Y_i$) and all edges in $F_i$, i.e. all edges $e$ with 
$2^i\leq \lambda_e\leq 2^{i+1}-1$. Let $G_0 = F_0$.
For any $i\not= j$, $F_i\cap F_j = Y_i\cap Y_j = \emptyset$; thus,
each edge appears in $G_i$ for at most two different values of 
$i$, proving $\alpha$-overlap. On the other hand, for any edge
$e\in F_i$, Lemma~\ref{lma:ni-con} ensures that the endpoints
of $e$ are connected in each of 
$T_{2^{i-1}}, T_{2^{i-1}+1}, \ldots, T_{2^i-1}$. It follows that
$e$ is $2^{i-1}$-heavy in $G_i$, thereby proving 
$\pi$-connectivity. 

We now prove the size bound on the skeleton.
Partition edges into subsets $S_0, S_1, \ldots$ where $S_j$ contains
all edges $e$ with $j < \frac{i_e}{w_e}\leq j+1$. The following lemma states
that none of these subsets is large.
\begin{lemma}
\label{lma:partition-ratio}
For any $j$, $|S_j|\leq n-1$.
\end{lemma}
\begin{proof}
We prove that the edges in any subset $S_j$ form an acyclic
graph. Suppose not; let $C$ be a cycle formed by the edge in 
$S_j$, and $e = (u, v)$ be the edge that was inserted last
in the NI forests among the edges in $C$. Let $e'$ be any 
other edge in $C$. Then, $w_{e'}\geq w_e$, and hence 
\begin{equation*}
i_{e'}+w_{e'}-1 > w_{e'}(j+1)-1 \geq w_e(j+1)-1 \geq i_e-1.
\end{equation*}
Since both the first and last terms are integers, 
$i_{e'}+w_{e'}-1 \geq i_e$. Therefore, $u'$ and $v'$ were 
connected in $T_{i_e}$ for each $e' = (u', v')$ in $C$. So,
$u$ and $v$ were connected in $T_{i_e}$ since $C$ is a cycle, 
before $e$ was added to $T_{i_e}$. But, then $e$ would not
have been added to $T_{i_e}$, a contradiction.
\end{proof}
\noindent
Thus, 
\begin{equation*}
\sum_e \frac{w_e}{i_e}\leq \sum_{j: S_j\not=\emptyset} \frac{|S_j|}{j}\leq
(n-1)\sum_{j: S_j\not=\emptyset} \frac{1}{j} = O(n\log n)
\end{equation*}
since at most $m < n^2$ of the $S_j$'s are non-empty. Using
Theorem~\ref{thm:main}, we conclude that the 
skeleton $G_{\epsilon}$ has $O(\frac{n\log^2 n}{\epsilon^2})$
edges in expectation and that $G_{\epsilon}\in (1\pm\epsilon)G$
whp.

Finally, we need to show that the construction of NI forests where
edges are added in decreasing order of weight can be done in 
$O(m\log^2 n)$ time. We use a data structure 
(call it a {\em partition tree}) $\cal P$ to succinctly encode 
the NI forests. The leaf nodes in $\cal P$ exactly correspond
to the vertices in graph $G$, i.e. there is a one-one mapping
between these two sets. On the
other hand, each non-leaf node $v$  of the partition tree
has a number $n(v)$ associated with it that satisfies the
following property: {\em for any two vertices $x, y$ in the 
graph, if $z$ be the {\em least common ancestor}\footnote{The
{\em least common ancestor} or lca of two nodes $x,y$ in a tree is 
the deepest node that is an ancestor of both $x$ and $y$.} 
of their corresponding leaf nodes in {\cal P}, then $x$ and $y$ are
connected in exactly the first $n(z)$ NI forests}. 
Then, 
$n(z)+1$ is the index of the first NI forest where edge $(x, y)$
is to be inserted. 
Initially, all the $n$ leaf nodes in $\cal P$ representing 
the graph vertices are children of the root node $r$, and
$n(r) = 0$. As edges are inserted in the NI forests, the 
partition tree evolves, but we make sure that the above 
property holds throughout the construction. 
Additionally, we also maintain the invariant that if $x$ is a 
child of $y$ in $\cal P$, then $n(x) > n(y)$. 

We need to show that we can maintain the above
properties of the partition tree as it evolves, and also
retrieve the lca of any pair of vertices efficiently for
this evolving partition tree. Let $(x, y)$ be 
the edge being inserted, let $z = lca(x, y)$ in 
the partition tree, and let $u$ and $v$ be the 
children of $z$ that are ancestors of $x$ and $y$
respectively. Observe that adding
an edge $(x, y)$ to trees with indices from 
$n_s+1$ to $n_s+\ell$ increases the connectivity of 
a pair of vertices $w_1, w_2$ iff they were previously 
connected in $n_s+i$ trees for some $0\leq i < \ell$, 
$w_1, x$ were connected in $n_s+j$ trees for some 
$j \geq i$ and $w_2, y$ were connected in $n_s+k$ trees 
for some $k \geq i$ (or vice-versa). In this case, $w_1, w_2$
are now connected in $n_s+\min(j, k, \ell)$ trees
after adding the edge $(x, y)$. Further,
if $n(u)-n(z) < w(x, y)$, then an edge of weight
less than $w(x, y)$ must have been added to the trees 
according to the second invariant, which violates
the fact that edges are added in decreasing order 
of weight. Thus, $n(u)-n(z) \geq w(x, y)$; similarly
$n(v)-n(z) \geq w(x, y)$.

There are three cases:
\begin{enumerate}
\item $n(u)-n(z) = n(v)-n(z) = w(x, y)$.
We merge $u$ and $v$ into a single node $s$ that
remains a child of $z$ and $n(s) = n(u)$.
The first invariant is clearly maintained. For
the second invariant, observe that the only 
pairs of vertices $w_1, w_2$ whose connectivity 
changed were those with $lca(w_1, w_2) = z$, where
$w_1, w_2$ are descendants of $u, v$ 
respectively. Their connectivity increases to
$n(u)$, which is reflected in the partition tree.
\item $n(u)-n(z) = w(x, y)$ and $n(v)-n(z) > w(x, y)$
(symmetrically for $n(u)-n(z) > w(x, y)$ and 
$n(v)-n(z) = w(x, y)$). 
We make $v$ a child of $u$ (from being a child of $z$),
and $n(u) = n(z)+w(x, y)$. 
For notational convenience in the proofs later, we 
replace $u$ and $v$ by a pair of new nodes $s$ and $t$ 
where $n(s)$ and $n(t)$ are respectively equal
to the updated values of $n(u)$ and $n(v)$.
The first invariant is clearly maintained. For
the second invariant, observe that the only 
pairs of vertices $w_1, w_2$ whose connectivity 
changed were those with $lca(w_1, w_2) = z$, where
$w_1, w_2$ are descendants of $u, v$ 
respectively. Their connectivity increases to
$n(z)+w(x, y)$, which is reflected in the partition tree.
\item $n(u)-n(z) > w(x, y)$ and $n(v)-n(z) > w(x, y)$.
We introduce a new node $r$ as a child of $z$ and
parent of $u$ and $v$, and $n(r) = n(z)+w(x, y)$.
For notational convenience in the proofs later, we 
replace $u$ and $v$ by a pair of new nodes $s$ and $t$ 
where $n(s) = n(u)$ and $n(t) = n(v)$.
The first invariant is clearly maintained. For
the second invariant, observe that the only 
pairs of vertices $w_1, w_2$ whose connectivity 
changed were those with $lca(w_1, w_2) = z$, where
$w_1, w_2$ are descendants of $u, v$ 
respectively. Their connectivity increases to
$n(z)+w(x, y)$, which is reflected in the partition tree.
\end{enumerate}

We use the {\em dynamic tree} data structure~\cite{SleatorT83} 
for updating the partition tree. This data structure can be
used to maintain a dynamically changing forest of $n$ nodes,  
while supporting the following operations\footnote{The dynamic 
tree data structure supports other operations as well; we only
define the operations that we require.} in $O(\log n)$ time 
per operation:
\begin{description}
\item[{\bf Cut}($v$)] Cut the subtree under node $v$ from 
the tree containing it, and make it a separate tree with root
$v$.
\item[{\bf Link}($v, w$)] ($w$ needs to be the root node of a 
tree not containing $v$.) Join the tree rooted at $w$ and that 
containing $v$ by making $w$ a child of $v$.
\item[{\bf LCA}($v, w$)] ($v$ and $w$ need to be in the same 
tree.) Defined previously.
\end{description}
We maintain a dynamic tree data structure for the partition
tree. Recall that the partition tree can be modified in three
different ways. The last two modifications require $O(1)$ cut 
and link operations each. Therefore, the overall time complexity
of these modifications is $O(m\log n)$. On the other hand, 
the first modification requires $O(d)$ cut and link operations, 
where $d$ is the lesser number of children among $u$ and $v$. 
We will prove the following lemma bounding the total number
of operations due to the first type of modification.
\begin{lemma}
\label{lma:mod-1}
The total number of cut and link operations due to modifications 
of the first type in the partition tree is $O(m\log n)$.
\end{lemma}
\noindent
Theorem~\ref{thm:weighted-tc} follows immediately.
\begin{theorem}
\label{thm:weighted-tc}
The time complexity of constructing NI forests where edges are
inserted in decreasing order of weight is $O(m\log^2 n)$ for
graphs with arbitrary edge weights.
\end{theorem}
\noindent
We now prove Lemma~\ref{lma:mod-1}.
\begin{proof}[Proof of Lemma~\ref{lma:mod-1}]
We set up a charging argument for the cut and link operations
due to the first type of modification. 
Define a function $f$ on the nodes of the partition tree where
each node $v$ has $f(v) = 1$ initially. In the first type of 
modification, we assign $f(s) = f(u)+f(v)$; in the second 
type of modification, $f(s) = f(u)+f(v)$ and 
$f(t)= 1$; in the third type of modification, 
$f(r) = f(u)+f(v)$ and $f(s) = f(t) = 1$.
Observe that the sum of $f(\cdot)$ over all nodes in the
partition tree increases by at most 2 for any of the
above modifications.

Let $C_u$ be the set of children of node $u$; then,
let $F_C(u) = \sum_{v\in C_u} f(v)$.
We charge the cut and link operations for the first type
of modification to the children of $u$ (resp., $v$) if
$F_C(u)\geq F_C(v)$ (resp., $F_C(v) > F_C(u)$); each child of 
$u$ (resp., $v$) is charged $O(1)$ operations. Now, let
$S_u$ be the set of siblings of any node $u$ in the partition
tree; correspondingly, let $F_S(u) = \sum_{v\in S_u} f(v)$.
Observe that whenever a node $u$ is charged due the first type 
of modification, $F_S(u)$ at least doubles. Further, $F_S(u)$
never decreases for any node $u$ due to any of the three types
of modifications. Since the sum of $f(.)$ over all nodes in the
partition tree increases by at most 2 for any of the modifications,
and there are $m$ modifications overall, each node is charged 
at most $O(\log m) = O(\log n)$ times. Further, each modification
introduces $O(1)$ new nodes; so the total number of operations
due to modifications of the first type is $O(m\log n)$.
\end{proof}

\bibliographystyle{plain}
\bibliography{ref}

\appendix

\section{Proof of Theorem~\ref{thm:compression}}
We need the following inequality.
\begin{lemma}
\label{lma:ineq-1}
Let $f(x) = x - (1+x)\ln (1+x)$ and $\alpha = 1- 2\ln 2$.
Then, 
\begin{equation*}
f(x) \leq
\begin{cases}
\alpha x^2 & {\rm if~} x\in (0, 1)\\
\alpha x & {\rm if~} x \geq 1.
\end{cases}
\end{equation*}
\end{lemma}
\begin{proof}
First, consider $x\in (0, 1)$. Define 
\begin{equation*}
g(x) = \frac{f(x)}{x^2} = \frac{1}{x} - \left(\frac{1}{x}+\frac{1}{x^2}\right)\ln (1+x).
\end{equation*}
We can verify that $g(x)$ is an increasing function of $x$ for 
$x \in (0, 1]$. Further, at $x = 1$, $g(x) = \alpha$. Thus,
$f(x) < \alpha x^2$ for $x\in (0, 1)$.

Now, consider $x\geq 1$. Define 
\begin{equation*}
h(x) = \frac{f(x)}{x} = 1 - \left(1+\frac{1}{x}\right)\ln (1+x).
\end{equation*}
We can verify that $h(x)$ is a decreasing function of $x$ for 
$x\geq 1$. Further, at $x = 1$, $h(x) = \alpha$. Thus,
$f(x) \leq \alpha x$ for $x\geq 1$.
\end{proof}
\noindent
We use the above inequality to prove the following lemmas.
\begin{lemma}
\label{lma:chernoff-extended-1}
Suppose $X_1, X_2, \ldots, X_n$ is a set of independent random
variables such that each $X_i$, $i\in \{1, 2,\ldots, n\}$, has 
value $1/p_i$ with probability $p_i$ for some fixed
$0 < p_i\leq 1$ and has value 0 with probability $1-p_i$. 
For any $p\leq \min_i p_i$ and for any $\epsilon > 0$,
\begin{equation*}
\pr\left[\sum_i X_i > (1 + \epsilon) n\right] < 
\begin{cases}
e^{-0.38\epsilon^2 pn} & {\rm if~} 0 < \epsilon < 1\\
e^{-0.38\epsilon pn} & {\rm if~} \epsilon \geq 1.
\end{cases}
\end{equation*}
\end{lemma}
\begin{proof}
For any $t > 0$,\footnote{For any random variable $X$, 
$\ex[X]$ denotes the expectation of $X$.}
\begin{eqnarray*}
\pr\left[\sum_i X_i > (1 + \epsilon) n\right] & = & \pr\left[e^{t\sum_i X_i} > e^{t(1 + \epsilon)n}\right] \\
& < & \frac{\ex\left[e^{t\sum_i X_i}\right]}{e^{t(1 + \epsilon)n}}\quad{\rm (by~Markov~bound~(see~e.g.~\cite{MotwaniR97}))} \\
& = & \prod_{i=1}^n \frac{\ex\left[e^{t X_i}\right]}{e^{t(1 + \epsilon)n}}\quad({\rm by~independence~of~} X_1, X_2, \ldots, X_n) \\
& = & \prod_{i=1}^n \frac{p_i e^{t/p_i} + 1 - p_i}{e^{t(1 + \epsilon)n}} \\
& = & \prod_{i=1}^n \frac{1 + p_i (e^{t/p_i}  - 1)}{e^{t(1 + \epsilon)n}} \\
& \leq & \exp(\sum_{i=1}^n p_i (e^{t/p_i}  - 1) - t(1 + \epsilon)n)\quad({\rm since~} 1 + x\leq e^x,~\forall x \geq 0).
\end{eqnarray*}
Since $p_i \geq p$ for all $i\in \{1, 2, \ldots, n\}$,
\begin{equation*}
\sum_{i=1}^n (p_i (e^{t/p_i}  - 1)) \leq \sum_{i=1}^n (p (e^{t/p}  - 1)) = np (e^{t/p}  - 1).
\end{equation*} 
Thus, 
\begin{equation*}
\pr\left[\sum_i X_i > (1 + \epsilon) n\right] < \exp(np (e^{t/p}  - 1) - t(1 + \epsilon)n).
\end{equation*}
Setting $t = p\ln (1+\epsilon)$, we get
\begin{equation*}
\pr\left[\sum_i X_i > (1 + \epsilon) n\right] < \left(\frac{e^\epsilon}{(1+\epsilon)^{1+\epsilon}}\right)^{pn}.
\end{equation*}
Since $1 - 2\ln 2 < -0.38$, we can use 
Lemma~\ref{lma:ineq-1} to conclude that
\begin{equation*}
\pr\left[\sum_i X_i > (1 + \epsilon) n\right] < 
\begin{cases}
e^{-0.38\epsilon^2 pn} & {\rm if~} 0 < \epsilon < 1\\
e^{-0.38\epsilon pn} & {\rm if~} \epsilon \geq 1.
\end{cases}\qedhere
\end{equation*}
\end{proof}

\begin{lemma}
\label{lma:chernoff-extended-2}
Suppose $X_1, X_2, \ldots, X_n$ is a set of independent random
variables such that each $X_i$, $i\in \{1, 2,\ldots, n\}$, has 
value $1/p_i$ with probability $p_i$ for some fixed
$0 < p_i\leq 1$ and has value 0 with probability $1-p_i$. 
For any $p\leq \min_i p_i$ and for any $\epsilon > 0$,
\begin{equation*}
\pr\left[\sum_i X_i < (1 - \epsilon) n\right] 
\begin{cases} 
< e^{-0.5 \epsilon^2pn} & {\rm if~} 0 < \epsilon < 1\\
= 0 & {\rm if~} \epsilon \geq 1.
\end{cases}
\end{equation*}
\end{lemma}
\begin{proof}
For $\epsilon \geq 1$,
\begin{equation*}
\pr\left[\sum_i X_i < (1 - \epsilon) n\right] \leq \pr\left[\sum_i X_i < 0\right] = 0.
\end{equation*}
Now, suppose $\epsilon \in (0, 1)$. For any $t > 0$, 
\begin{eqnarray*}
\pr\left[\sum_i X_i < (1 - \epsilon) n\right] & = & \pr\left[e^{-t\sum_i X_i} > e^{-t(1 - \epsilon)n}\right] \\
& < & \frac{\ex\left[e^{-t\sum_i X_i}\right]}{e^{-t(1 - \epsilon)n}}\quad{\rm (by~Markov~bound)} \\
& = & \prod_{i=1}^n \frac{\ex\left[e^{-t X_i}\right]}{e^{-t(1 - \epsilon)n}}\quad({\rm by~independence~of~} X_1, X_2, \ldots, X_n) \\
& = & \prod_{i=1}^n \frac{p_i e^{-t/p_i} + 1 - p_i}{e^{-t(1 - \epsilon)n}} \\
& = & \prod_{i=1}^n \frac{1 - p_i (1 - e^{-t/p_i})}{e^{-t(1 - \epsilon)n}} \\
& \leq & \exp(\sum_{i=1}^n -p_i (e^{-t/p_i}  - 1) + t(1 - \epsilon)n)\quad({\rm since~} 1 - x\leq e^{-x},~\forall x \geq 0).
\end{eqnarray*}
Since $p_i \geq p$ for all $i\in \{1, 2, \ldots, n\}$,
\begin{equation*}
\sum_{i=1}^n (p_i (1 - e^{-t/p_i})) \leq \sum_{i=1}^n (p (1 - e^{-t/p})) = np (1 - e^{-t/p}).
\end{equation*} 
Thus, 
\begin{equation*}
\pr\left[\sum_i X_i < (1 - \epsilon) n\right] < \exp(np (1 - e^{-t/p}) + t(1 - \epsilon)n).
\end{equation*}
Setting $t = -p\ln (1-\epsilon)$, we get
\begin{equation*}
\pr\left[\sum_i X_i < (1 - \epsilon) n\right] 
< \left(\frac{e^\epsilon}{(1-\epsilon)^{1-\epsilon}}\right)^{pn}
\leq e^{-0.5 \epsilon^2pn}.\qedhere
\end{equation*}
\end{proof}
\noindent
We now prove Theorem~\ref{thm:compression} using the 
above lemmas.
\begin{proof}[Proof of Theorem~\ref{thm:compression}]
Let $\delta = \frac{\epsilon N}{|C|}$. 
First, consider the case where $\delta \in (0, 1)$.
From Lemmas~\ref{lma:chernoff-extended-1} and 
\ref{lma:chernoff-extended-2}, we conclude that
\begin{eqnarray*}
\pr\left[|\sum_e X_e - |C|| > \epsilon |C|\right] & = & 
\pr\left[|\sum_e X_e - |C|| > \delta |C|\right] < 
2 e^{-0.38\delta^2 p|C|}\\ 
& = & 2 e^{-0.38\epsilon^2 pN (N/|C|)} \leq
2 e^{-0.38\epsilon^2 pN}\quad({\rm since~} N\geq |C|).
\end{eqnarray*}

Now, consider the case where $\delta \geq 1$.
From Lemmas~\ref{lma:chernoff-extended-1} and 
\ref{lma:chernoff-extended-2}, we conclude that
\begin{equation*}
\pr\left[|\sum_e X_e - |C|| > \epsilon N\right] = 
\pr\left[|\sum_e X_e - |C|| > \delta |C|\right] < 
e^{-0.38\delta p|C|} =
e^{-0.38\epsilon pN} \leq
e^{-0.38\epsilon^2 pN}\quad({\rm since~} \epsilon\leq  1).\qedhere
\end{equation*}
\end{proof}

\end{document}